\documentclass[aps,prb,reprint,showpacs,superscriptaddress,byrevtex]{revtex4-2}
\usepackage[normalem]{ulem}
\usepackage{graphicx}
\usepackage{dcolumn}
\usepackage{bm}
\usepackage{physics}
\usepackage{svg}

\usepackage{hyperref}
\usepackage[mathlines]{lineno}

\begin{document}

\preprint{APS/123-QED}

\title{Shift photoconductivity in the Haldane model}

\author{Javier Sivianes}
\affiliation{Centro de F\'{i}sica de Materiales (CSIC-UPV/EHU), 20018, Donostia-San Sebasti\'{a}n, Spain}
\author{Julen Iba\~{n}ez-Azpiroz}%
\affiliation{Centro de F\'{i}sica de Materiales (CSIC-UPV/EHU), 20018, Donostia-San Sebasti\'{a}n, Spain}
\affiliation{Ikerbasque Foundation, 48013 Bilbao, Spain}
\affiliation{Donostia International Physics Center (DIPC), 20018
Donostia-San Sebasti\'{a}n, Spain}

\date{October 18, 2023}

\begin{abstract}
The shift current is part of the second-order optical response of materials with a close connection to topology. Here we report a sign inversion in the band-edge shift photoconductivity of the Haldane model when the system undergoes a topological phase transition. This result is obtained following two complementary schemes. On one hand, we derive an analytical expression for the band-edge shift current in a two-band tight-binding model showing that the sign reversal is driven by the mass term. On the other hand, we perform a numerical evaluation on a continuum version of the Haldane model. This approach allows us to include off-diagonal matrix elements of the position operator, which are discarded in tight-binding models but can contribute significantly to the shift current. Explicit evaluation of the shift current shows that while the model predictions remain accurate in the deep tight-binding regime, significant deviations arise for shallow potential landscapes. Notably, the sign reversal across the topological phase transition is observed in all regimes, implying it is a robust effect that could be observable in a wide range of topological insulators. 
\end{abstract}


\maketitle

\section{Introduction}

Over the last decades, interest has grown on the second order optical response known as the bulk photovoltaic effect (BPVE)~\cite{fridkin_bulk_2001,sturman-book92,ivchenko-book97}. This is partly due to its prospects for the development of efficient solar cells not bound by the Shockey-Queisser limit \cite{Shockley,Spanier2016,tan-cm16}. For a material subject to an electrical field $\mathbf{E}(\omega)$, the BPVE generates a DC current $\mathbf{j}$ that can be written as
\begin{equation}
 j^{a}(0)=\sigma^{abc}(0;\omega,-\omega)E^{b}(\omega)E^{c}(-\omega),\label{eq:current}
\end{equation}
where superscripts refer to Cartesian components and $\sigma^{abc}(0;\omega,-\omega)$ is the photoconductivity tensor, which we denote simply as $\sigma^{abc}(\omega)$ from now on. Given the quadratic nature of the BPVE, it can only be present in noncentrosymmetric materials \cite{VonBaltz}.


Multiple contributions to the BPVE exist, such as the shift and 
injection currents; these can arise from linearly or circularly polarized light, where the polarization controls how the relevant matrix elements are combined~\cite{Zhang2019, Wang2020}. In particular,
the linear shift current has attracted considerable attention lately \cite{1PhysRevResearch.4.013164,2PhysRevB.104.115402,3doi:10.1021/acs.jpclett.0c03503,PhysRevB.103.245415,PhysRevResearch.4.013164,tan_effect_2019, PhysRevLett.125.227401, PhysRevB.108.075413}, and its photoconductivity tensor is given by~\cite{Sipe}
\begin{eqnarray}\label{eq:shift}
    \sigma^{abc}(\omega) = &&-\frac{i\pi{e}^3}{2\hbar^2}\int \frac{d\mathbf{k}}{(2\pi)^3}\sum_{mn}f_{nm}I_{mn}^{abb} \\ \nonumber
    &&\times\left[\delta\left(\omega_{nm}-\omega\right)+\delta\left(\omega_{mn}-\omega\right)\right].
\end{eqnarray}
Here $f_{nm}=f_{n}-f_{m}$ is the occupation factor difference, $\omega_{mn}=\omega_{m}-\omega{n}$ is the energy gap of the bands involved and the transition matrix element 
\begin{equation}\label{eq:Imn}
    I_{mn}^{abc}=r_{mn}^{b}r_{nm}^{c;a}     
\end{equation}
contains the dipole term $r_{mn}^{b}=i(1-\delta_{nm})\bra{n}\partial_{a}\ket{m}$ and its generalized derivative $r_{nm}^{c;a}=\partial_a r_{nm}^{c}-i\left(A_{nn}^{a}-A_{mm}^{a}\right)r_{nm}^{c}$ with $A_{nn}^{a}=i\bra{n}\partial_{a}\ket{n}$ the intraband Berry connection. 
In this context, the shift current can be interpreted as the real-space shift of electrons upon an interband transition, encoded in the shift vector $R_{nm}^a = \partial_a \phi_{nm} + A_{nn}^a-A_{mm}^a$ with $r_{mn}=\abs{r_{mn}}e^{-i\phi_{mn}}$ \cite{Sipe}. 

Recently, the geometric interpretation of non-linear optical responses has lead to a connection to the field of topology \cite{Nagaosa2017-kc,topnature, PhysRevB.107.155434}. In the shift current, the geometry of the wave function comes into play through the Berry connection $A_{nn}$, which describes the relation between bordering wave functions belonging to the manifold with band index $n$. As a consequence of the role of geometric aspects, the shift current is enhanced in materials like Weyl semimetals \cite{Ma2019,deJuan2017,PhysRevB.102.121111,PhysRevB.107.205204, PhysRevB.100.155201, Osterhoudt2019}. Another particularly appealing effect was reported in the work of Tan and Rappe \cite{tan2016enhancement}, where DFT calculations in $\text{BiTeI}$ and $\text{CsPbI}_{3}$ revealed that the shift current undergoes a sign change across the topological phase transition (TPT). This singular behavior could potentially be exploited for the experimental determination of topological states by purely optical means. Given its potential practical interest, it becomes desirable to gain insight into the fundamental aspects of this peculiar effect.

Considering the above, the well-known Haldane model represents an ideal test system for various reasons. First, it describes a Chern insulator hosting a TPT between a trivial and a topological insulator~\cite{haldane1988model}. Secondly, it allows inversion symmetry breaking, a mandatory requirement for a non-zero shift current. Finally, thanks to its relatively simple two-band model structure, the resulting analysis offers a clear description of the sign reversal effect and its regime of validity.  

In this work, we examine the band-edge shift current in the Haldane model by means of two complementary schemes. First, based on a $\mathbf{k}\cdot\mathbf{p}$ expression for the transition matrix element \cite{cook2017design}, we show that the low-energy current reverses its sign upon the TPT due to a sign change of the mass term. Additionally, we show that in the topological insulating phase the shift photoconductivity tensor at the two valleys in the Brillouin zone (BZ) has opposite signs, resulting in a sharp discontinuous jump in the secondary gap. In order to evaluate the generality of the model results, in a second step we conduct an exact numerical evaluation of the shift current in a continuum system that maps into the Haldane model. Our calculations show that, while the model predictions remain valid in the deep-tight-binding regime, significant quantitative and qualitative deviations arise for shallow potential landscapes that mimic known two-dimensional (2D) materials. Notably, the band-edge sign reversal through the TPT is observed in all regimes despite the numerical differences, suggesting that the effect is very robust and largely independent of the potential describing the system.

This paper is organized as follows. In Sec. \ref{sec:twoband} we present our analytical calculation of the shift current on the Haldane model and discuss the role of the TPT and the $\mathbf{k}\rightarrow-\mathbf{k}$ asymmetry in the band structure. Then, in Sec. \ref{sec:numerical} we describe the results of the numerical evaluation and analyze the quantitative importance of off-diagonal matrix elements of the position operator, not included in the tight-binding model. Finally, we provide conclusions and outlook in Sec. \ref{sec:conclusions} and describe TPT's between higher-order Chern numbers in Appendix.

\section{Shift current in the Haldane model}\label{sec:twoband}
\subsection{Review of the model}\label{sec:model}

The Haldane model is built on a 2-D honeycomb lattice formed by a triangular lattice containing a two atom basis \cite{haldane1988model}. This results in two sublattices which we refer to as A and B. The relevant vectors involved are; the lattice vectors $\mathbf{a}_1=a/2(\sqrt{3},3)$ and $\mathbf{a}_2=3a/2(\sqrt{3},-3)$,  the shift between nearest neighbors (nn) $\mathbf{e}_1=(0,a)$ and $\mathbf{e}_{2,3}=a/2(\mp \sqrt{3},-1)$, and the displacement between next-nearest neighbors (nnn) $\mathbf{v}_1=(-\sqrt{3}a,0)$ and $\mathbf{v}_{2,3}=a/2(\sqrt{3},\mp 3)$, where $a$ is the lattice parameter. The properties of the system are controlled by a tight-binding Hamiltonian with the following elements:
\begin{itemize}
    \item The on-site energies +M(-M) for the A(B) site.
    \item The nn hopping parameter $t_1$.
    \item The nnn hopping parameter $t_2e^{\pm i\phi}$ with +(-) for the A(B)$\rightarrow$A(B) term.
\end{itemize}

The appearance of the phase $\phi$ in the nnn hopping parameter is due to the inclusion of a vector potential with the periodicity of the cell under the condition that the net magnetic flux is null. As a loop on the full lattice can be performed by hopping along nn, $t_{1}$ does not pick up a phase, while $t_{2}$ does for the opposite reason. This leads to the breaking of the time reversal symmetry.

The Hamiltonian expressed in $k$-space takes the form \cite{pires2019brief}

\begin{eqnarray}
\mathcal{H}( \mathbf{k}) && =\sum_i f_{i}(\mathbf{k})\sigma_{i}\label{eq:H},\\ \nonumber
f_{0} && =2t_{2}\cos\phi\left[\sum_{i=1}^{3}\cos(\mathbf{k}\cdot \mathbf{v_{i}})\right],\\ \nonumber,
f_{1} && =t_{1}\sum_{i}\cos({\mathbf{k}\cdot \mathbf{e_{i}}}),\\\nonumber
f_{2} && =t_{1}\sum_{i}\sin({\mathbf{k}\cdot \mathbf{e_{i}}}),\\\nonumber
f_{3} && =M-2t_{2}\sin\phi\sum_{i}\sin(\mathbf{k}\cdot \mathbf{v_{i}}).
\end{eqnarray}

The band structure of the model is given by $\epsilon=\sqrt{\sum_i f_i}$ and the band-edges are located at the corners of the BZ, which are the $\mathbf{K}=(\frac{4\pi}{3\sqrt{3}a},0)$ and $\mathbf{K'} = (-\frac{4\pi}{3\sqrt{3}a},0)$ points. The band gap $\Delta(\mathbf{k})$ at the two valleys can be explicitly written as 
\begin{equation}
    \Delta=2\left(M+\chi3\sqrt{3}t_{2}\sin\phi\right)=2 m(\chi), \label{eq:gap} 
\end{equation}
where $\chi = -1(+1)$ makes reference to the K(K$'$) point, and $m(\chi)$ is known as the mass term. Eq. (\ref{eq:gap}) shows that the gap at the corners of the BZ is a function of the tight-binding parameters. Note that for $t_2\sin\phi\neq 0$ their values differ. Whenever $m(\chi)<0$, a band inversion takes place, implying a change in the topology of the system. For the Haldane model, the Chern number characterizes the topological order of the system, which can be written as
\begin{equation}
    C=\frac{1}{4\pi}\int_{BZ}\frac{\mathbf f}{\norm{f}^3}\left(\frac{\partial\mathbf f}{\partial k_{x}}\times\frac{\partial\mathbf f}{\partial k_{y}} \right)dk_{x}dk_{y}.
\end{equation}
There are three different scenarios within the Haldane model; if $\abs{M} = 3\sqrt{3}\abs{t_{2}\sin\phi}$ the system is gapless, $\abs{M} > 3\sqrt{3}\abs{t_2\sin\phi}$ represents a trivial insulator $(C=0)$ and $\abs{M} < 3\sqrt{3}\abs{t_2\sin\phi}$ describes a topological insulator $(C=\pm 1)$ \cite{el-batanouny_2020}. This implies that the lowest energy band-edge gap changes sign upon the TPT. Another important observation for the proceeding section is that $m(\chi)$ acquires a different sign at the two valleys only in the topological insulator phase.

\subsection{Band-edge shift current: sign change at the TPT}\label{sec:analytic}
In order to derive an analytical expression for the shift current, we make use of a simplified formulation for the transition matrix elements in Eq. (\ref{eq:shift}) that was given in Ref. \cite{cook2017design} and is valid for two-band models:
\begin{equation}
I_{12}^{abb}=\sum_{ijm}\frac{1}{4\epsilon^3}\left(f_{m}f_{i,b}f_{j,ab}-f_{m}f_{i,b}f_{j,a}\frac{\epsilon_{,b}}{\epsilon}\right)\epsilon_{ijm},\label{eq:I12}
\end{equation}
with $f_{i,a}=\partial_af_i$ and $\epsilon_{ijm}$ the Levi-Civita tensor.

Given that we are interested in the properties at the band-edges, we perform a second order Taylor expansion of Eq. (\ref{eq:H}) around the K and K$'$ points. Due to the presence of a second-order derivative in Eq. (\ref{eq:I12}), we need to Taylor-expand up to second order in momentum. The resulting low-energy Hamiltonian is 
\begin{eqnarray}
f_{1} && =\frac{3}{2}t_{1}\left(\chi q_{x}a-\frac{1}{4}q_{y}^{2}a^{2}+\frac{1}{4}q_{x}^{2}a^{2}\right)\\
f_{2} && =\frac{3}{2}t_{1}\left(q_{y}a-\frac{1}{2}\chi q_{x}q_{y}a^{2}\right)\\
f_{3} && = m(\chi) + \chi\frac{9\sqrt{3}}{4}t_{2}\sin(\phi)\left(q_{x}^2a^2+q_{y}^2a^{2}\right).
\end{eqnarray}
Here $m(\chi)$ is the mass term as defined in Eq. (\ref{eq:gap}) and $q_{i}\,(i=x,y)$ are the momenta centered at the corners of the BZ. The $f_{0}$ term need not be specified since it does not affect the transition matrix elements in Eq. (\ref{eq:I12}). For energies of the incident radiation close to the band-edge $\hbar\omega\approx2\abs{m(\chi)}$, Eq. (\ref{eq:shift}) can be factorized as \cite{cook2017design}
\begin{equation}
    \sigma^{abb}(\omega) = -\frac{\pi\text{e}^3}{4\hbar^2} I_{12}(\omega) N(\omega). \label{eq:factoriz}
\end{equation}
where $N(\omega) = \int \frac{dq^2}{(2\pi)^2}\delta(2\epsilon-\hbar\omega)$ is the joint density of states (JDOS) and $2\epsilon$ is the energy difference between the two bands \cite{cook2017design}.

The evaluation of Eq. (\ref{eq:I12}) results in 
\begin{eqnarray}\label{eq:model-I}
I_{12}^{yyy}(\omega) && =\frac{9a^{3}t_{1}^{2}\text{sign}[m(\chi)]}{8\left(\hbar\omega\right)^2}+\mathcal{O}(q)\label{eq:Iyyy}\\
I_{12}^{xyy}(\omega) && = -\frac{9a^{3}t_{1}^{2}\text{sign}[m(\chi)]}{8\left(\hbar\omega\right)^2}+\mathcal{O}(q).\label{eq:-12}
\end{eqnarray}
As expected for the $C_{3v}$ point group of the Haldane model, $\sigma^{yyy} = -\sigma^{yxx}$, while the rest of components vanish by symmetry.

Next, in order to evaluate $N(\omega)$ at the band-edge we perform a Taylor expansion in the energy difference of the bands and express it in polar coordinates as
\begin{eqnarray}
\hbar\omega_{12}(q)\approx2\abs{m(\chi)}+\frac{\hbar^2q^2}{M'},
\end{eqnarray}
where $\frac{\hbar^2}{M'}=\frac{9a^2t_{1}^2-18\sqrt{3}a^2mt_{2}\chi\sin(\phi)}{4|m(\chi)|}$. Then, the JDOS simplifies to
\begin{equation}
N (\omega)={\frac{\hbar \omega \,\Theta\left(\hbar\omega-2\abs{m(\chi)}\right)}{18\pi a^{2}t_{1}^{2}\abs{\eta(\chi)}}}\label{eq:JDOS}
\end{equation}
with $\eta(\chi)=1-2\sqrt{3}\abs{m(\chi)}t_{2}\sin(\phi)/t_{1}^2$.

Note that for $t_{2}\neq0$, the JDOS for the expanded Hamiltonian at  K and {K$^{\prime}$} points is not equal due to the $\eta(\chi)$ term. 

Plugging  Eqs. (\ref{eq:Iyyy}) and (\ref{eq:JDOS}) into Eq. (\ref{eq:factoriz}), the  shift current tensor at the band edge reads:
\begin{equation}\label{eq:sigmayyy}
\sigma^{yyy}(\omega) =-\frac{e^{3}a}{4\hbar^3\omega\abs{\eta\left(\chi\right)}}\text{sign}\left[m(\chi)\right] = -\sigma^{yxx}(\omega)
\end{equation}

One important aspect of the above expressions is that the tensor components depend on the sign of the mass term $m(\chi)$, which changes across the TPT. Eq. (\ref{eq:sigmayyy}) therefore reproduces the sign reversal effect found by DFT calculations in topological insulators $\text{BiTeI}$ and $\text{CsPbI}_{3}$ \cite{tan2016enhancement}, and is also in accordance with the general conclusions of a related work~\cite{https://doi.org/10.48550/arxiv.1812.02191}. 
In the Appendix \ref{sec:appendix} we extend these 
results and show that 
the sign reversal of the band-edge shift current also
holds for TPT's involving higher-order Chern numbers.
As a second major point of Eq. (\ref{eq:sigmayyy}), in the topological phase the photoconductivity contributions at the two valleys have opposite sign owing to the valley-dependence of the mass term in Eq. (\ref{eq:gap}). 
Therefore, an additional sign reversal can arise at the largest band-gap value of the two valleys.
In the coming section we study in more detail these two particular features.

\section{Testing model predictions}\label{sec:numerical}

 In order to complement the analysis of the previous section, we now consider a continuum version of the Haldane model and compute the shift current exactly following the numerical scheme of Ref. \cite{WannierInterpolation}. The purpose of this procedure is to verify to what extent does the simplified two-band model expression in Eq. (\ref{eq:sigmayyy}) describe the actual shift current response. In general, this is composed by both Hamiltonian and position matrix elements \cite{WannierInterpolation}. While tight-binding models can account for the matrix structure of the Hamiltonian via tunneling terms between different sites, the position is implicitly diagonal~\cite{bennetto-prb96,PhysRevB.66.165212,lee-prb18}. To study this in greater detail, let us express the position operator in a localized Wannier basis $\ket{\textbf{R}m}$ \cite{RevModPhys.84.1419} as
 \begin{equation}
\bra{\mathbf{0}n}\mathbf{r}\ket{\mathbf{R}m} = \bm{\tau}_m\delta_{\mathbf{0}\mathbf{R}}\delta_{nm}+\mathbf{d}_{nm}(\mathbf{R}). \label{eq:off}
\end{equation}
Here $\bm{\tau}_m\delta_{\mathbf{0}\mathbf{R}}\delta_{nm}$ corresponds to the onsite or diagonal element of the position operator, while $\mathbf{d}_{nm}$ are the hopping or off-diagonal elements of the position operator. In tight-binding models $\mathbf{d}_{nm}=0$ generally, so the contribution of these terms is not included in Eq. (\ref{eq:sigmayyy}). 
Refs.~\cite{pedersen-prb01,PhysRevB.72.125105} have assessed
the impact of  intra-atomic $\mathbf{d}_{nm}$
in the linear optical response of toy models.
More recent works have shown that the effect of $\mathbf{d}_{nm}$ on the shift current can be of the order or the Hamiltonian matrix elements \cite{wang-prb17,WannierInterpolation}, and that it becomes specially large for two-band systems \cite{ibanez2022assessing}. Therefore, it is reasonable to ask if the predictions encoded into the two-band model expressions of Eq. (\ref{eq:sigmayyy}) hold in reality. This is our main purpose in the remainder of this work.
\begin{figure}[t]
    \includegraphics[width = 0.48 \textwidth]{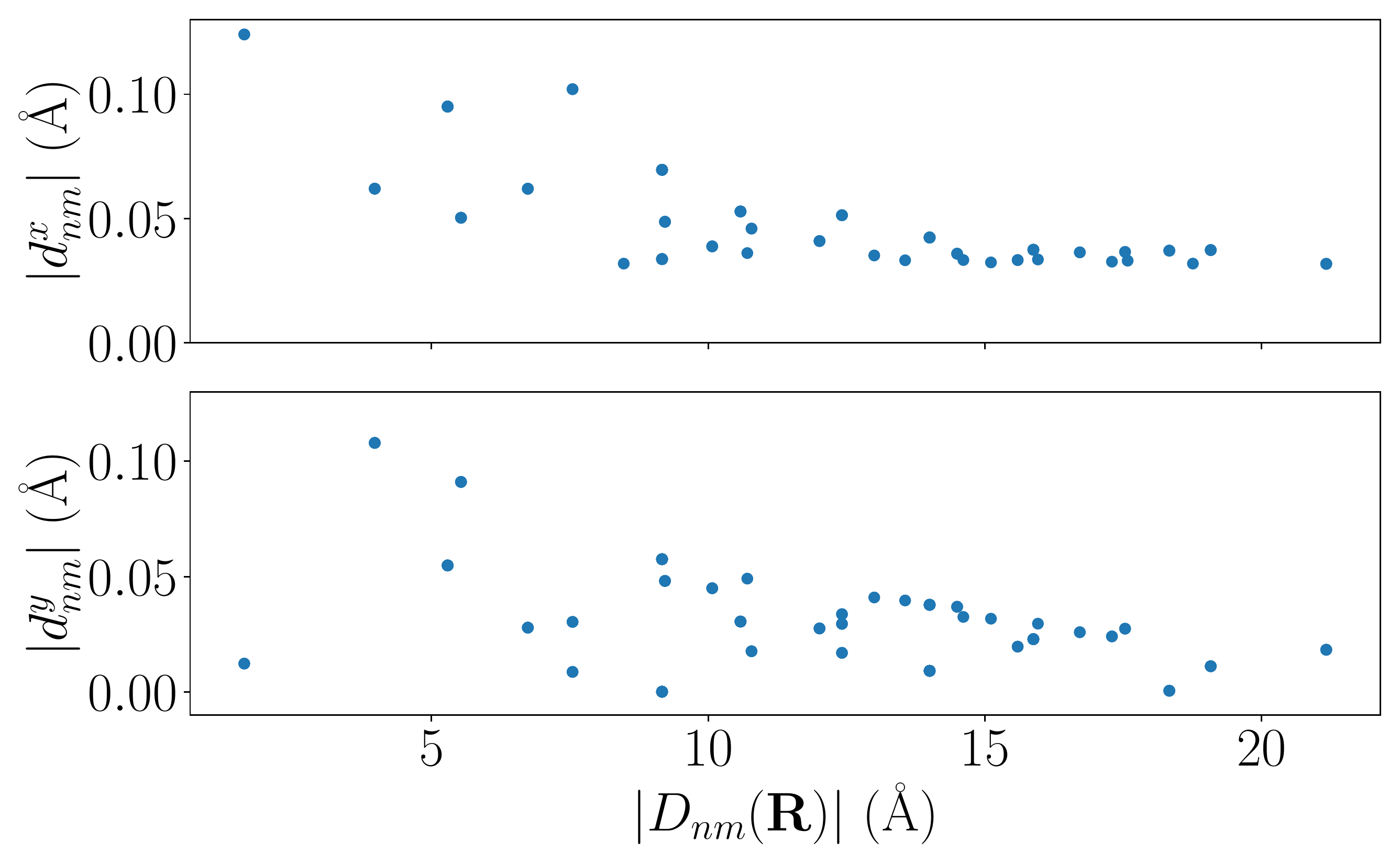}
    \caption{Cartesian components of off-diagonal matrix elements of the position operator as a function of the distance between Wannier orbitals. Employed parameters are $s=0.70$, $\psi=1.5\cdot10^{-3}$ and $\alpha=0.27$.}
    \label{fig:D}
\end{figure}

 \subsection{Continuum model}
 
 We consider the continuum model of Ref. \cite{shao2008realizing}, which was proposed to mimic the Haldane model in the context of cold atoms trapped in optical lattices. Here the lattice is spanned by the lattice vectors $\mathbf{a}_{1,2}= (2\pi/3k_L)(\mathbf{e}_x\mp\sqrt{3}\mathbf{e}_y)$ ($k_{L}$ is the so-called laser wave vector) \cite{shao2008realizing}. Hence, the Hamiltonian 
 \begin{equation}
 \mathcal{H}=\frac{1}{2m}\left[\mathbf{p}-\mathbf{A}(\mathbf{r})\right]^{2}+V_{L}(\mathbf{r}) \label{eq:Hcont}   
 \end{equation}
 is characterized by a scalar potential $V_{L}$($\mathbf r$) and a vector potential $\mathbf A$($\mathbf r$). The former is responsible of generating the honeycomb lattice, and has the form 
\begin{eqnarray}
    V_L(\mathbf{r})=&&2 s E_{R}\Bigl\{\cos\left[\left(\mathbf{b}_1-\mathbf{b}_2\right)\cdot \mathbf{r}\right]+\cos\left(\mathbf{b}_1\cdot\mathbf{r}-\frac{\pi}{3}\psi\right)\nonumber\\ 
    &&+\cos\left(\mathbf{b}_2\cdot\mathbf{r}\right)\Bigr\}, \label{eq:V}
\end{eqnarray}
where $E_{R}=\hbar^{2}k_{L}^2/2m$ is the recoil energy, $s$ characterizes the amplitude of the potential in units of $E_{R}$, $\mathbf{b}_i$ are the reciprocal lattice vectors, and $\psi$ controls the inversion symmetry breaking. The vector potential $\mathbf A$($\mathbf r$), given by 
\begin{eqnarray}
    \mathbf{A}(\mathbf{r})=&&\alpha \hbar k_{L} \biggl[ \Bigl\{ \sin
    \left[\left(\mathbf{b}_2-\mathbf{b}_1\right)\cdot \mathbf{r}\right] +\frac{1}{2}\sum_{i=1}^{2}(-1)^i \nonumber \\ 
    &&\sin\left(\mathbf{b}_i\cdot\mathbf{r}\right) \Bigr\} \mathbf{e}_x-\frac{\sqrt{3}}{2}\sum_{i=1}^{2}\sin \left(\mathbf{b}_i\cdot \mathbf{r}\right)\mathbf{e}_y\biggr]
\end{eqnarray}
breaks time reversal symmetry for $\alpha\neq0$, which is the amplitude of the potential. In this continuum model, the tunnelings and on-site energies $t_{1}$, $t_{2}$ and $M$ depend on the parameters that control the potentials, namely $s$, $\psi$ and $\alpha$ \cite{haldanecontinuum}.

\subsubsection{Calculation details}

The procedure that we have employed for the numerical evaluation can be summarized in three steps. First, we solve for the eigenvalues and eigenfunctions of Hamiltonian (\ref{eq:Hcont}) following the approach described in Refs.~\cite{PhysRevA.87.011602,peierls}. Then, maximally localized Wannier functions are constructed from Bloch eigenfunctions using the software package Wannier90 \cite{Pizzi2020}. This is done by minimizing the spread of the Wannier functions
\begin{equation}
\Omega=\sum_{n}\left[\bra{\textbf{0}n}r^2\ket{\textbf{0}n}-\bra{\textbf{0}n}\textbf{r}\ket{\textbf{0}n}^2\right],\label{eq:spread}
\end{equation}
which is a measure of their degree of localization~\cite{marzari-prb97,souza-prb01}. Finally, we compute the shift current following the procedure outlined in Ref.~\cite{WannierInterpolation}.

Regarding the parameters involved in the simulations, we have generated the continuum model  following the procedure outlined in Refs.~\cite{haldanecontinuum,peierls}. We have a employed a 
15$\cross$15 $k$-point mesh as a basis to construct Wannier functions, while 
we have employed a fine  2000$\cross$2000 mesh through Wannier interpolation 
to compute the shift current in Eq. (\ref{eq:shift}). 

\begin{figure}[b]
     \includegraphics[width = 0.48\textwidth]{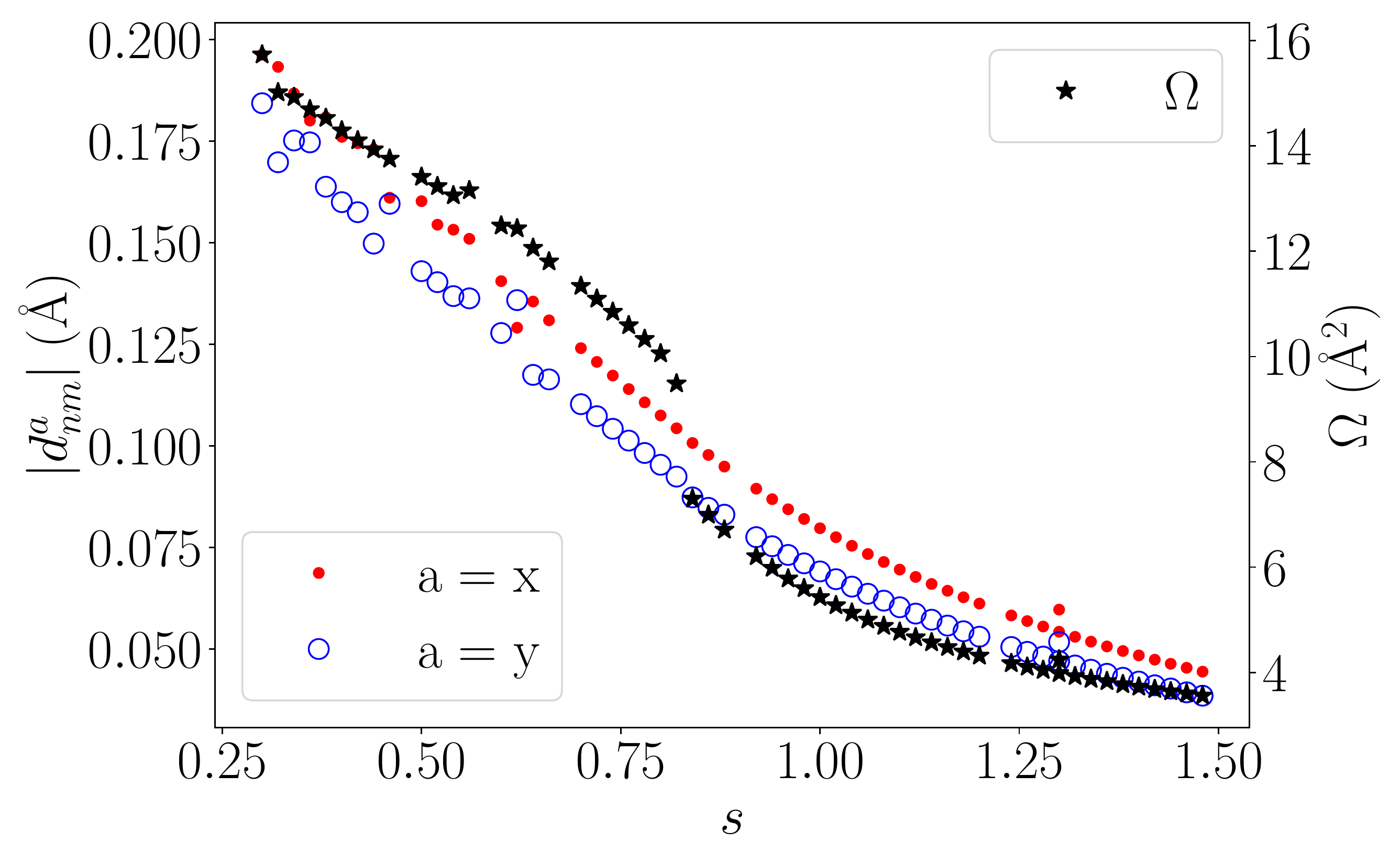}
     \caption{Cartesian components of the dominant off-diagonal matrix elements of the position operator as a function of $s$ and its correlation to the spread of the Wannier functions. Employed parameters are $\psi=1.5\cdot10^{-3}$ and $\alpha=0.27$.}
     \label{fig:dxdys}
 \end{figure}

\begin{figure*}
    \includegraphics[width = 0.48\textwidth]{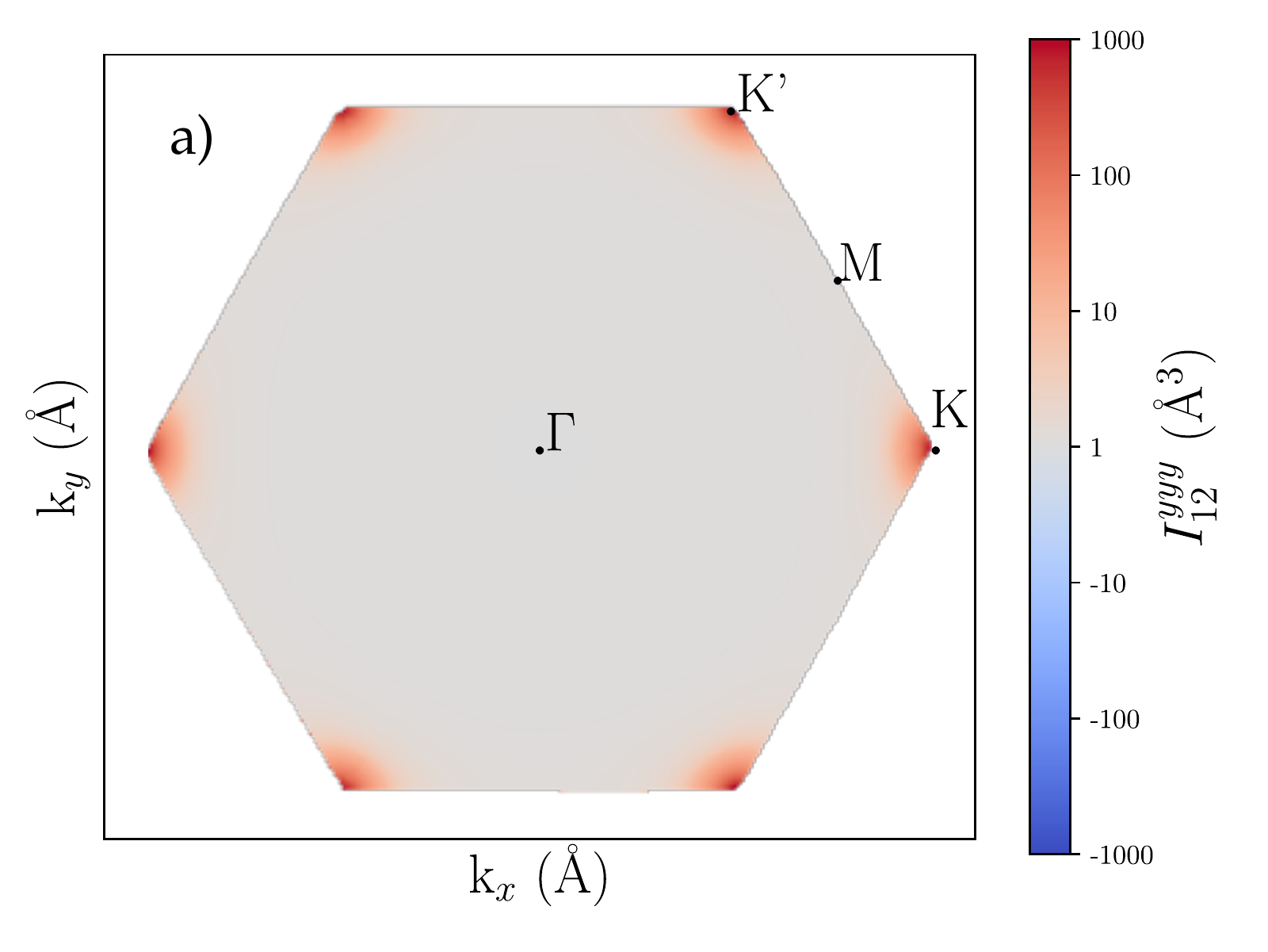}
    \hfill
    \includegraphics[width = 0.48\textwidth]{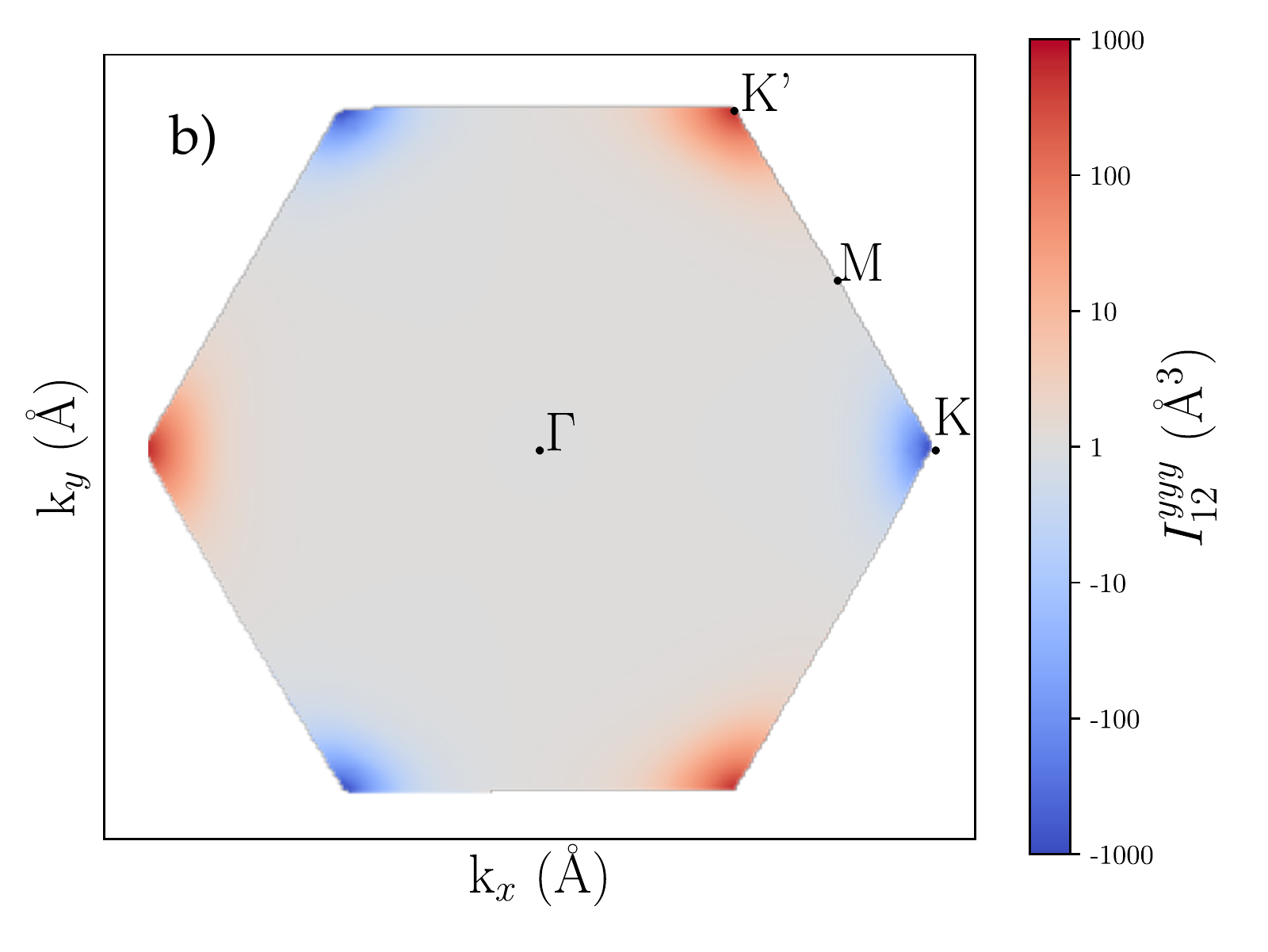}
    \caption{Representation of $I^{yyy}_{12}$ in the BZ for the (a) $C = 0$  and (b) $C = 1$ phases obtained with parameters $\alpha = 0.0 $, $\psi =1.5\cdot10^{-3}$, $s = 2.5$  and  $\alpha = 0.13 $, $\psi =1.5\cdot10^{-3}$, $s = 2.5$, respectively.}
\label{fig:kslice}
\end{figure*}

\subsection{Matrix elements of the position operator}\label{sec:offdiagonal}
 
We begin by studying the effect of off-diagonal position matrix elements ($\mathbf{d}_{nm}$ in Eq. \ref{eq:off}). For this purpose, let us first set the real-space dimensions of the system. In our implementation we have chosen $k_L = 0.78\ \text{\r{A}}^{-1}$, leading to the lattice vectors $\mathbf{a}_{1,2} = (2.65, \mp 4.58)\ \text{\r{A}}$, lattice parameter $a=3.05\ \text{\r{A}}$ and Wannier centers $\mathbf{\tau}_{1,2} = (1.40, \mp0.81)\ \text{\r{A}}$. In this way, our system has similar dimensions to real 2D monolayers \cite{2dlattice} such as BC$_2$N studied in Ref. \cite{ibanez2022assessing}, which will serve  as a reference.

Fig. \ref{fig:D} presents the Cartesian components of $\mathbf{d}_{nm}$ as a function of the real-space distance between orbitals
\begin{equation}
    D_{nm}(\mathbf{R})=\abs{\bm{\tau}_{n}-\bm{\tau}_{m}+\mathbf{R}}.
\end{equation}
The results have been calculated for a setup that is inversion asymmetric with $\psi=1.5\cdot10^{-3}$, TR-broken with $\alpha=0.27$ and relatively shallow potential well of $s=0.70$.

We observe that the predominant contribution to $d^x_{nm}$ comes from the nn within the unit cell at $D_{nm} \sim 1.6\ \text{\r{A}}$, but neighbors as far as 10 $\text{\r{A}}$ still contribute with half the maximum value. For $d^y_{nm}$ the dominant contribution is the nnn term at  $D_{nm} \sim 4.0\ \text{\r{A}}$, and the decay with distance is similar to that of $d^x_{nm}$. We note that these trends are in line with the two-band position matrix elements as well as the magnitudes of monolayer BC$_2$N \cite{ibanez2022assessing}.
 
Next, we analyze the dominant contributions of $d_{nm}^x$ and $d_{nm}^y$ for varying $s$, represented in Fig. \ref{fig:dxdys}. In addition, we also plot the spread of the Wannier functions in Eq. (\ref{eq:spread}) as a function of $s$. Our results show that the $\textbf{d}_{nm}$ increase significantly in the low $s$ regime, which correlates with a large spread $\Omega$. Therefore, the model parameter $s$ effectively allows to control the importance of off-diagonal position matrix elements. As pointed out earlier, the magnitude of the calculated $\mathbf{d}_{nm}$ for low $s$ is of the same order as the one found in real materials such as monolayer BC$_2$N \cite{ibanez2022assessing}. 

\subsection{Numerical shift current}
\subsubsection{Deep tight-binding regime}

\begin{figure}
\includegraphics[width = 0.48\textwidth]{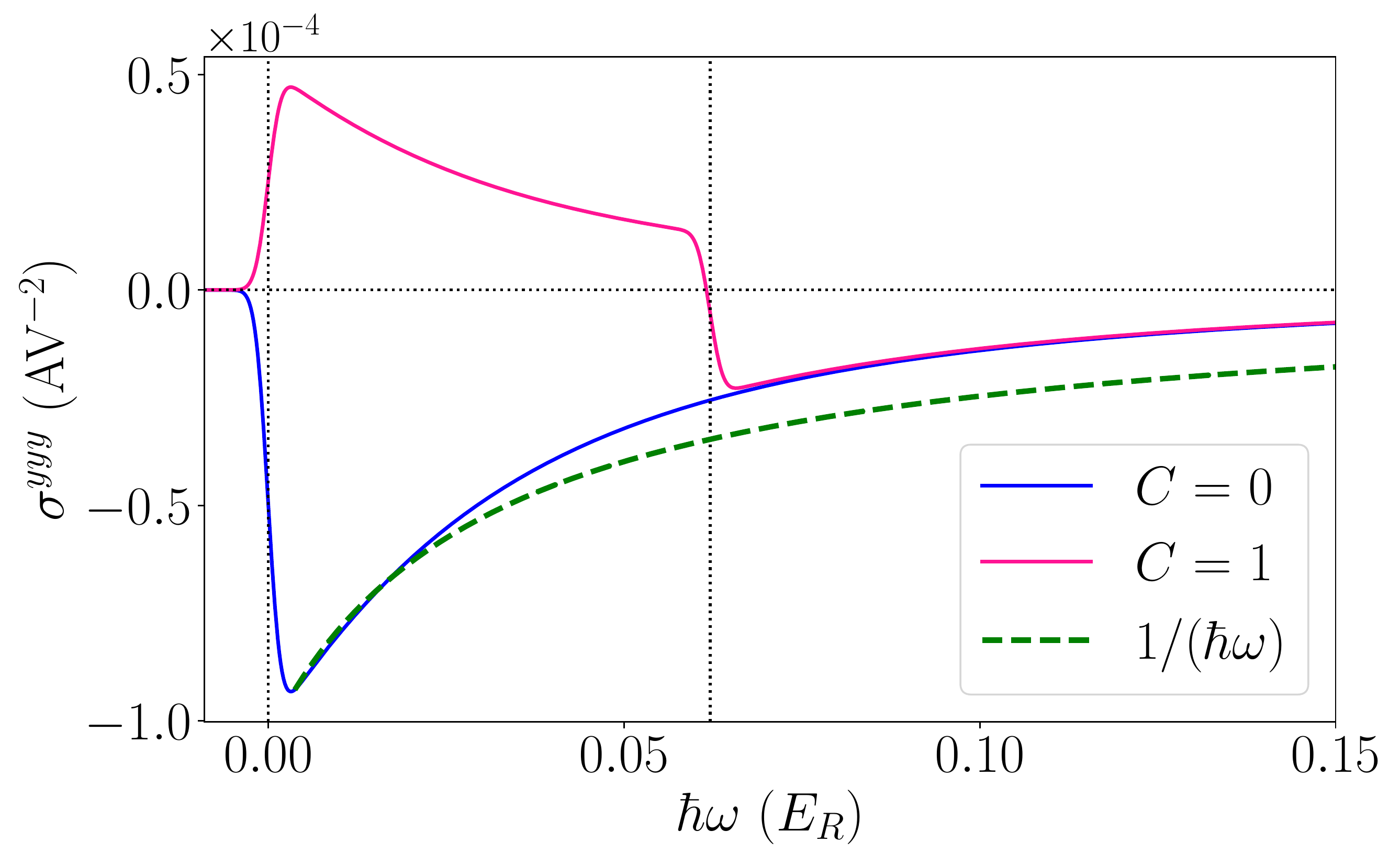}
\caption{\label{fig:sc} Shift photoconductivity $\sigma^{yyy}(\omega)$ of the trivial (blue) and topological (magenta) systems described in Fig. \ref{fig:kslice}. The energy is measured with respect to each ones first band-gap and vertical lines mark the gaps at the band-edges of the topological insulator. A $1/\hbar\omega$ decay
for the $C=0$ case is also included (green dashed line).}
\end{figure}

We come now to test the results of the tight-binding analysis. For this purpose, we start analyzing the deep tight-binding regime (large $s$) and choose two sets of parameters that describe a Haldane model in its trivial and non-trivial insulating phases \cite{haldanecontinuum}. 
The associated transition matrix element $I_{12}^{yyy}$ is shown in Fig. \ref{fig:kslice}. As revealed by the figure, $I_{12}^{yyy}$ is highly localized in the two valleys, which justifies the $\mathbf{k}\cdot\mathbf{p}$ expansion performed in Sec. \ref{sec:analytic}. We note that the calculated sign at K and K$'$ agrees with the model predictions encoded in Eq. \ref{eq:Iyyy}
for both the $C=0$ and $C=1$ cases.

We next focus on analyzing the resulting shift current tensor $\sigma^{yyy}(\omega)$, shown in Fig. \ref{fig:sc} for both phases. We observe that the responses associated to the lowest energy band-edge transition have opposite signs for the two phases (in this example the gap at both valleys for the $C =0$ phase are equal). Furthermore, the shift photoconductivity at the second
band-gap also manifests a sign inversion in the $C=1$ case, which is due to  the opposite sign of the transition 
matrix element at the two valleys [see Fig.~\ref{fig:kslice}(b)].  
The characteristics described are in accordance with Eq. (\ref{eq:sigmayyy}), revealing that the two-band shift current expression appropriately describes the main response features close to the band-edge. 

Concerning the behavior of the photoconductivity around the band-edge, in Fig.~\ref{fig:sc} we have included for the $C=0$ case the $1/\omega$ decay predicted by the band-edge model expression of Eq. (\ref{eq:sigmayyy}). As shown in the figure, the fit reproduces fairly well the decay close to the band-edge but underestimates it at higher energies.


\subsubsection{Limits of the TB description}

\begin{figure}
    
    \includegraphics[width = 0.48\textwidth]{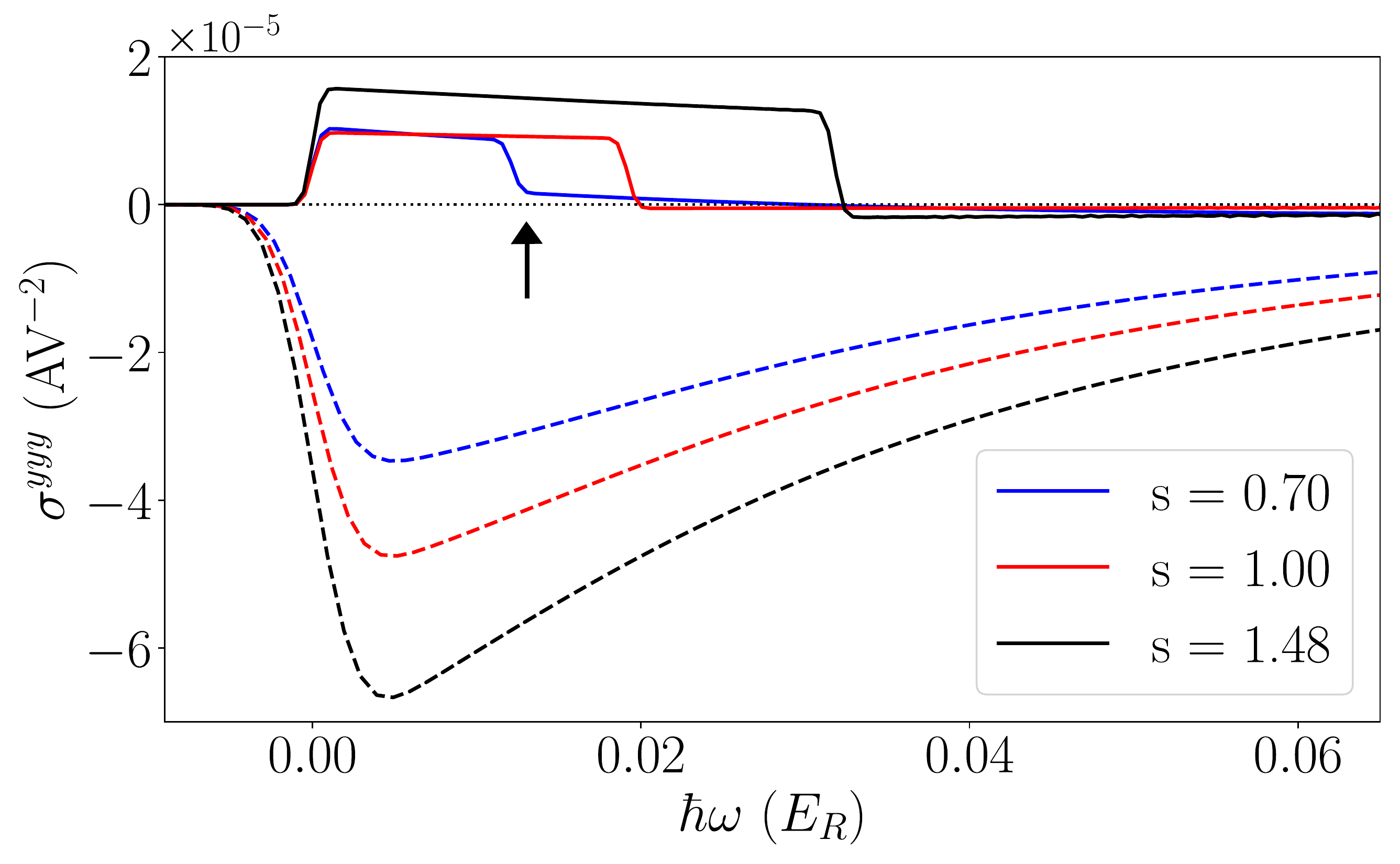}
    \caption{Shift photoconductivity $\sigma^{yyy}(\omega)$
    in the continuum Haldane model for three values of $s$. For each value of $s$ a trivial insulator (dashed) and a topological insulator (solid) is represented, with parameters $\alpha = 0.0$, $\psi = 1.5\cdot10^{-3}$ and $\alpha = 0.27$, $\psi = 1.5\cdot10^{-3}$, respectively. For $s = 0.70$, the sign at the second band-edge is not reversed, a feature that is marked by the arrow.}
    \label{fig:merge}
\end{figure}

In order to verify the extent of validity of the model predictions, we next compute the shift current as a function of the parameter $s$. This is illustrated in Fig. \ref{fig:merge}, where we show the calculated $\sigma^{yyy}$ for three values of $s$. As observed in the figure, the sign reversal at the lowest band-gap across the TPT is maintained along the entire range of $s$. 
This implies that it is a robust effect that is present beyond the 
deep tight-binding regime. 
However, that is not the case for the sign inversion at the second band-gap in the topological phase, which fails to take place for $s = 0.70$ (see marking arrow in Fig.~\ref{fig:merge}).

\begin{figure}
    \includegraphics[width = 0.48\textwidth]{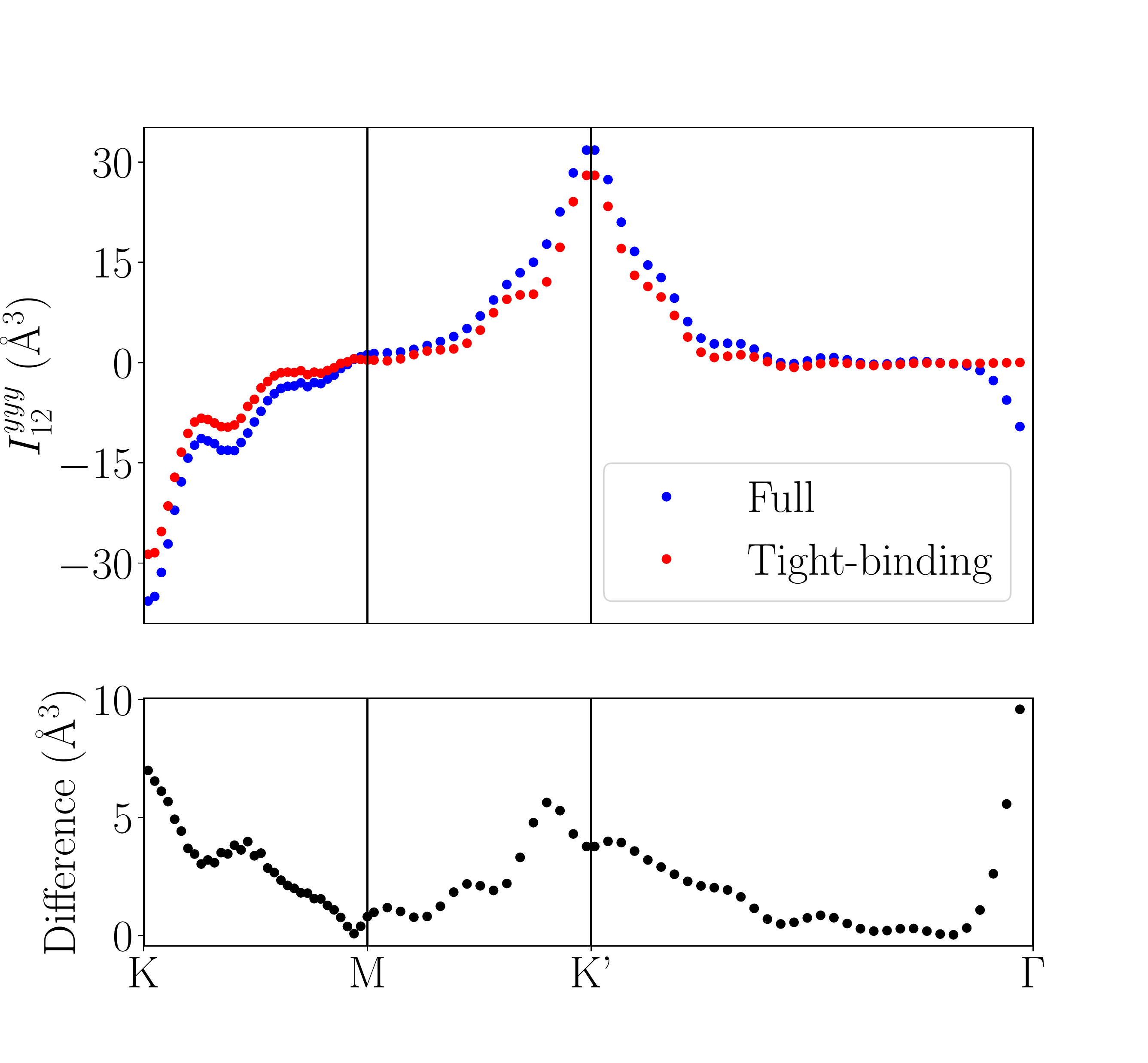}
    \caption{ Top: tight-binding and full matrix element $I_{12}^{yyy}$ along the high symmetry line K-M-K$^{\prime}$-$\Gamma$. Bottom: the difference between the two data sets, which accounts for the contribution from $\mathbf{d}_{mn}$ to $I_{12}^{yyy}$.}
    \label{fig:path}
\end{figure}

To inspect the latter feature in more detail,  in Fig. \ref{fig:path} 
we show the transition matrix element $I_{12}^{yyy}$ of Eq. (\ref{eq:shift}) along the 
high symmetry line K-M-K$^{\prime}$-$\Gamma$ for the $C=1$ case with $s = 0.70$. 
Apart from the exact calculation of $I_{12}^{yyy}$ we have also included
a ``tight-binding-like'' contribution obtained by setting $\mathbf{d}_{nm}=0$,
which serves to identify the extent of their contribution. 
The maximum difference between the two sets takes place at $\Gamma$ but it is also 
very pronounced at the K and K$^{\prime}$ points. 
Importantly, the effect of the
$\mathbf{d}_{nm}$ is inequivalent at the two valleys,
leading to $|I_{12}^{yyy}(\text{K})|=35.7\ \text{\r{A}}$ 
and $|I_{12}^{yyy}(\text{K}^{\prime})|=31.8\ \text{\r{A}}$.
In contrast, according to the model 
expression of Eq.(\ref{eq:Iyyy}) it should have  virtually the same value 
given that the band-gap at the valleys has similar energy 
$\Delta (\text{K}^{\prime})$=$1.02 \Delta (\text{K})$;
this is actually the case for the ``tight-binding'' transition matrix element, 
with $|I_{12}^{yyy}(\text{K})|=28.68 \ \text{\r{A}}$ 
and $|I_{12}^{yyy}(\text{K}^{\prime})|=28.01\ \text{\r{A}}$. 
It is precisely the fact that $|I_{12}^{yyy}(\text{K})|>|I_{12}^{yyy}(\text{K}^{\prime})|$
in the full calculation that prevents the shift photoconductivity to flip 
sign at the gap in {K} marked by the arrow in Fig. \ref{fig:merge}. 
This circumstance neatly exemplifies the importance of off-diagonal position matrix elements, which can modify not only quantitatively but also qualitatively the model predictions for the shift photoconductivity.


\section{Conclusions and outlook}\label{sec:conclusions}

In summary, we have obtained a two-band tight-binding expression for the nonlinear
shift photoconductivity in the  Haldane model. It describes an  optical sign inversion when the system undergoes a topological phase transition and is driven by the mass term. Additionally, the model expression predicts a further sign change of the shift photoconductivity in the topological phase at the band-gap of the second valley. 
In a subsequent step, we have assessed the extent of validity of these properties based on exact numerical evaluation of the shift current on a continuum version of the model. This approach incorporates off-diagonal matrix elements of the position operator $\mathbf{d}_{nm}$, which are not included in the tight-binding approach. We have found that the contribution of $\mathbf{d}_{nm}$ to the shift current is significant  for shallow potential landscapes that mimic known 2D materials. While the model predictions for the secondary sign change fail far from the tight-binding regime, the main band-edge sign change takes place in all inspected regimes of the continuum model. 

The above suggests that the sign reversal of the shift current across the
topological phase transition is a robust effect that might therefore be experimentally observed in 
real topological insulators.
The pool of materials include two-dimensional 
quantum anomalous Hall systems 
such as thin films of Cr-doped (BiSe)$_2$Te$_3$~\cite{doi:10.1126/science.1234414}
and interfaces between ferromagnetic and non-magnetic semiconductors~\cite{Chang_2019}
(see, \textit{e.g.}, Ref.~\cite{doi:10.1080/00018732.2015.1068524} for more examples).
In these systems, the topological phase transition 
of interest
can be induced by means of an external magnetic field, which plays the role of the complex phase $\phi$ in the Haldane model. 
Additionally, the conclusions of the present work 
can also be valuable in 
three-dimensional topological insulators such as 
$\text{BiTeI}$ and $\text{CsPbI}_{3}$,
given that their topology described by the Kane-Mele model \cite{PhysRevLett.95.226801}
is driven by the inversion of the mass term at the band-edge, a feature 
that is shared with the Haldane model.
In these classes of topological materials, the topological phase transition 
can be induced by applying pressure and effectively tuning the lattice parameter,
as described in the work of Tan and Rappe~\cite{tan2016enhancement}.

\begin{acknowledgments}
We are very grateful to Ivo Souza and Fernando de Juan for helpful discussions in the early stages of this work. This project has received funding from the European Union’s Horizon 2020 research and innovation programme under the European Research Council (ERC) Grant Agreement No. 946629,  and the Department of Education, Universities and Research 
of the Eusko Jaurlaritza and the University of the 
Basque Country UPV/EHU (Grant No. IT1527-22).
\end{acknowledgments}

\appendix

\section{Higher Chern number phase transitions}\label{sec:appendix}

In this appendix we expand on the results obtained in Sec. \ref{sec:analytic} by 
considering a topological phase transition between non-trivial phases. Within the Haldane model, a topological phase with $\abs{C}=2$ can be achieved 
by including the third nearest neighbor tunneling
$t_{3}$ into the model, as proposed in Ref.~\cite{PhysRevB.87.115402}. 
This then opens the possibility for phase transitions between
the $C=\pm1$ and $C=\mp2$ phases (see phase diagram in Ref.~\cite{PhysRevB.87.115402}).
In the following we study the shift current under such phase transitions.

Since the third nearest neighbor connects A sites with B sites, 
it results in an extra term in the Hamiltonian coefficients that go with the Pauli matrices $\sigma_1$ and $\sigma_2$:
\begin{eqnarray}
    f'_1  =& f_1 + t_3 \sum_{i}^{3}\cos(\mathbf{k}\cdot\mathbf{c}_i)\\
    f'_2 =& f_2 + t_3 \sum_{i}^{3}\sin(\mathbf{k}\cdot\mathbf{c}_i)
\end{eqnarray}
Here $\mathbf{c}_i$ are the vectors connecting to third nearest neighbors~\cite{PhysRevB.87.115402}.
An important difference with the standard Haldane model 
is that the band-closing point does not take place
exactly at the $K$ (or $K^\prime$) point 
but at three points around it. 
One of such points is given by
\begin{eqnarray}
    \Tilde{K}^{(\prime)} = K^{(\prime)} +\chi \delta k(1,0),
\end{eqnarray}
with $\delta k$ a small displacement;  
the remaining two gap-closing points are found by applying the $C_3$ symmetry operation.

Expanding the Hamiltonian around $\Tilde{K}^{(\prime)}$ results in lengthy expressions for the coefficients $f^{\prime}_i$. 
The most important change for our purpose takes place in the $f_3$ term  containing the mass term, which is modified as:
\begin{eqnarray}
     m'(\chi) = m(\chi) + \frac{9\sqrt{3}a^2 \delta k^2 t_2 \sin(\phi)}{4} + O(\delta k^3)
\end{eqnarray}
This modified mass term drives the topological phase transition as its sign is reversed. 

The new shift-current transition matrix element $I_{12}^{yyy}$ computed with the aid of Eq.(~\ref{eq:I12}) reads
\begin{eqnarray}
    I_{12}^{yyy} \simeq & \dfrac{a^3\left(9 t_1^2 +18 t_1 t_3 -36 t_3^2\right)}{8\hbar^2 \omega^2}\text{sign}[m'(\chi)]. \label{Iappendix}
\end{eqnarray}

As in Eq.~(\ref{eq:Iyyy}) for the standard Haldane model,  
Eq. (\ref{Iappendix}) above depends on the sign of the 
modified mass term.
In practice, this is due to the fact that the band-edge contribution 
to $I_{12}^{abb}$ is determined by the terms
$f_3 f_{1,b} f_{2,ab}$ and $f_3 f_{2,b}f_{1,ab}$ in Eq.~(\ref{eq:I12}), which contain the mass term only once. 
Therefore, Eq. (\ref{Iappendix}) shows that 
the band-edge shift current also undergoes a sign flip
at the TPT between the $C=\pm1$ and $C=\mp2$ non-trivial phases.    
This result agrees with the conclusion of Ref. \cite{https://doi.org/10.48550/arxiv.1812.02191}
derived on the basis of a more generic argument.

\color{black}


\bibliography{name}

\end{document}